\documentclass[aps,superscriptaddress,twocolumn]{revtex4}
\usepackage{amsmath,amssymb,graphics,graphicx,dcolumn,bm,enumerate}
\usepackage{mathbbol}

\newcommand{\tcmp}
{\affiliation{TCMP Division,
Saha Institute of Nuclear Physics, 1/AF Bidhannagar, Kolkata 700 064 India.}}
\newcommand{\imsc}
{\affiliation{Institute of Mathematical Sciences, CIT Campus, Taramani, Chennai-600113, India
}}

\begin{document}

\title{Equivalence of the train model of earthquake and boundary driven Edwards-Wilkinson interface}

\author{Soumyajyoti Biswas}
\email[Email: ]{soumyajyoti.biswas@saha.ac.in}
\tcmp

\author{Purusattam Ray}
\email[Email: ]{ray@imsc.res.in}
\imsc

\author{Bikas K. Chakrabarti}
\email[Email: ]{bikask.chakrabarti@saha.ac.in}
\tcmp 

\begin{abstract}
 \noindent  A discretized version of the Burridge-Knopoff train model with (non-linear friction force replaced by) 
random pinning is studied in one and two dimensions. A scale free distribution of avalanches and the Omori law type 
behaviour for after-shocks are obtained. The avalanche dynamics of this model becomes precisely similar (identical 
exponent values) to the Edwards-Wilkinson (EW) model of interface propagation. It also allows the complimentary 
observation of depinning velocity growth (with exponent value identical with that for EW model) in this train model 
and Omori law behaviour of after-shock (depinning) avalanches in the EW model.
\end{abstract}

%%%%%%%%%%%%%%%%%%%%%%%%%%%%%%%%%%%%%%%%%%%%%%%%%%%%%%%%%%%%%%%%%%%%%%%%%%%%%%%%%%%%%
\maketitle
\section{Introduction}
\label{sec:1}
\noindent The spring-block system driven over a rough surface has been studied extensively 
as a model for intermittent dynamics seen in earthquakes \cite{bk} (see \cite{rmp2} for a review). 
With a non-linear velocity dependent friction force and
linear springs, these models are known to produce rich dynamics. In particular, a well known statistical
law of earthquakes, such as the Guttenberg-Richter law, which predicts a scale free probability 
distribution for the sizes of avalanches, is reproduced in this simple model. 

One version of the spring-block model is the train model \cite{vieira}, where the blocks are 
connected by springs and the entire system glides over a rough surface as one block (the engine) 
at one extreme end is pulled. This version is much simpler (to analyze and simulate) than the   
Burridge-Knopoff spring-block model \cite{bk}, where the blocks are connected by springs to a 
moving plane which drags the blocks over a rough surface. In these models, the rough surface 
introduces friction and the blocks move to the positions of zero force once they  
overcome the static frictional force (threshold). Movement of one block induces forces on the 
neighboring connected blocks and this triggers movements of 
other blocks and the process of stick-slip motion of blocks continues. Though the model 
reproduces some features of earthquakes, the presence of velocity-dependent 
frictional force  makes analysis and simulation difficult. Further simplification is achieved in 
the train model \cite{chianca} by assuming a static friction that is a 
stochastic function of position of the blocks instead of being velocity dependent.
However, 
introduction of static friction depending on the position and the asperities of the surfaces and blocks at 
different positions render  the model complicated and difficult to analyze.

The spring-block model, in essence, captures the threshold activated processes 
giving rise to the intermittent dynamics. In our model, proposed here, we consider this exclusively 
and replace the frictional forces  by random pinning forces. Our model is a variant of the train model, 
where the blocks are placed at the sites of a lattice and connected by springs. In one (two) dimensions(s), 
two such chains (planes) of blocks 
go past each other as the upper chain (plane) is pulled at the boundary. 
When two blocks from upper and lower planes come on top of each other, a random pinning 
$F^{pin}$ is introduced. A block slides only when the total force on the block $F > F^{pin}$ and goes 
to the next lattice site. In our model, $F^{pin}$ is a random dynamical variable chosen between 0 and 1 every time 
a block slides and comes on top of another block.

We use a different definition for avalanches to capture both the Guttenberg-Richter law and the Omori 
law, which are the two prototypical laws concerning the statistical features of earthquake dynamics. 
It is to be noted that in the original Burridge-Knopoff model and in the train models, the Omori law could not 
be reproduced. We distinguish each of the slip events subsequent to the primary one and
consider each event occurring at subsequent time steps. This can be justified from the fact that
in earthquakes, the aftershocks are essentially a consequence of the primary adjustments (slips) of the 
existing contact points and not due to
further movement (and thereby stress building) of the tectonic plates, which moves in a time scale orders of 
magnitude slower than at least the initial aftershock event's time. Using this definition of avalanche, we get
both the well known statistical laws of earthquakes, namely the Guttenberg-Richter law and Omori law.

We simulate the model and use finite size scaling analysis to determine the exponent values associated  with 
the avalanche statistics. We suggest that the essential ingredient of the dynamics is pinning and randomness 
and show the universality between the models with random threshold and uniform lattice constant and 
uniform threshold and random (with substitutional defect) lattice constant. We have also studied the situation where the equilibrium 
spacing of the blocks follow a Cantor set (self affine but deterministic) and find that the exponent values do not change. 
We have extended our studies to two dimensions.
Finally, we show that our version of the train model is precisely equivalent to the interface propagation model of
Edwards-Wilkinson's (EW) \cite{ew} (see \cite{bouchaud} for detailed discussion on dynamical transitions on
interface models) pulled at the 
boundaries.

\section{Model}\label{sec:2}

\noindent Consider the one dimensional train model, where an array of blocks are arranged on a discrete 
lattice. The blocks are connected by Hookean springs (with identical spring constants). The array is
being pulled from one side quasistatically, i.e. the block on the extreme right say, is pulled until instability
sets in, after which the pulling is paused as long as there are movements of the blocks. Once all the movements
stop, the right most block is pulled again and so on. This puts the system in an intermittent motion after some 
transients. Three forces act on each block (except for the two blocks at the two extreme ends), two forces from 
two nearest neighbor springs and another due to friction on that site.  In our modified train model, 
the major simplification  
is made by replacing friction force by a random pinning force. The value of the friction force may depend on the
properties of the two surfaces in contact and therefore on every new contact, this value gets changed. In our
model, the pinning is a random variable chosen between 0 and 1 every time 
a block slides and come on top of another block. In a previous attempt \cite{chianca} to consider friction as random forces,
the roughness properties of the surfaces were explicitly considered in obtaining the friction force values. Here
we do not consider explicitly the roughness properties. 

The dynamics of the system was carried out as follows: If $r_i(t)$ denotes the position of the $i$-th block of the upper chain
at time $t$, then the net force on that block at that time is given by
\begin{equation}
F_i^{tot}(t)=F_i^{el}(t)+C(r_i(t))F_i^{pin},
\label{dyn1}
\end{equation}
where $F_i^{el}$ is the net elastic force due to the two neighbouring blocks and $F_i^{pin}$ is the pinning force, which acts against 
the direction of motion and $C(r_i(t))$ is the
condition if the block at position $r_i(t)$ of the upper chain has another block below it or not (so $C$ can take binary value 0 or 1).
The expression for the net elastic force is given by  
\begin{eqnarray}
F^{el}_{i}(t)=&&k(r_{i+1}(t)-r_i(t)-R^0_{i,i+1})\nonumber \\
&&+k(R^0_{i-1,i}-(r_{i}(t)-r_{i-1}(t)))
\label{dyn2}
\end{eqnarray}
where $R^0_{i,i+1}$ denotes the equilibrium spacing between $i$-th and $(i+1)$-th block.
Now if $F^{tot}_{i}>0$, then the position of the $i$-th block is moved by unit distance provided there is no block
 at that position. The extreme left block 
does not have a neighbour on one side, so the equation is to be adjusted accordingly and that on the extreme
right is the `engine'. 
%%%%%%%%%%%%%%%%%%%%%%%%%%%%%%%%%%%%%%%%%%%%%%%%%%%%%%%%%%%%%%%%%%%%%%%%%%%%%%%%%%%%%%%%%%%%%%%%%%%%%%%%%%%%%%%%%%%%%%%%%%%%%%%%%%%%%%%%%
\begin{figure}[tb]
\centering \includegraphics[width=8.2cm]{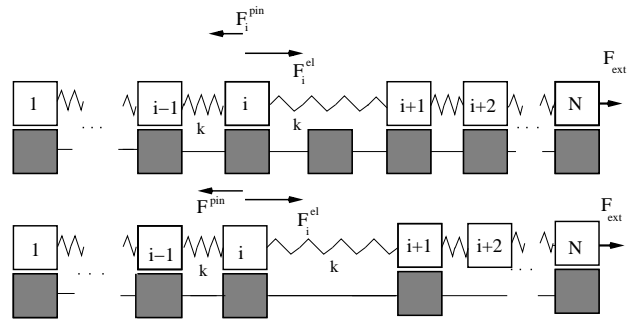}
   \caption{Schematic diagram for the discrete spring-block model with pinning. The upper figure depicts the case where there is
no disorder in the lattice spacings of the chains (IC) and the lower figure has disorder in the lattice spacing of the chains (DC).
The directions of elastic force and pinning force (working as friction) are shown for a typical block. The chains are being pulled from
one end quasistatically (see text). The pinning forces in the IC case are distributed in a range (typically [$0:1$]) but for the DC case,
pinning force has a single value; it only comes into play when two blocks come on top of each other and the randomness comes from the
randomness in the positions of the blocks in the both the chains.}
\label{chain}
\end{figure}
%%%%%%%%%%%%%%%%%%%%%%%%%%%%%%%%%%%%%%%%%%%%%%%%%%%%%%%%%%%%%%%%%%%%%%%%%%%%%%%%%%%%%%%%%%%%%%%%%%%%%%%%%%%%%%%%%%%%%%%%%%%%%%%%%%%%%%%%%%
%%%%%%%%%%%%%%%%%%%%%%%%%%%%%%%%%%%%%%%%%%%%%%%%%%%%%%%%%%%%%%%%%%%%%%%%%%%%%%%%%%%%%%%%%%%%%%%%%%%%%%%%%%%%%%%%%%%%%%%%%%%%%%%%%%%%%%%%%

We mainly consider two types of chains: One is like the usual train model, where the blocks are placed an equal
distance apart; let us call this the intact chain (IC). A non-zero friction (or pinning) force can only
appear when two blocks come on top of each other, otherwise the friction is zero. The pinning force is a 
random number drawn from the uniform distribution [$0:1$]. Remembering that the fault surfaces are rough, 
we also consider the case when some of the blocks are removed from both the upper and lower chains; let us call
this a disordered chain (DC). For this case the pinning force is always unity for all the blocks in the train
in contact with the block in the lower chain. As we shall see, even without a distribution in the friction
force, the system shows intermittent dynamics, which suggests that the effect of roughness of the surface on
dynamics can be incorporated into the disorders of the chains. Remembering that the rough surfaces are often 
self-affine, we also consider the case when the blocks in the two chains are arranged in the form of a Cantor set 
of same generation; let us call this self-affine chain (SAC). This differs from DC in the sense that the disorders here 
are correlated.   

\section{Results}

\subsection{Avalanche statistics in modified train model} 

\begin{figure}[tb]
\centering \includegraphics[height=6.7cm]{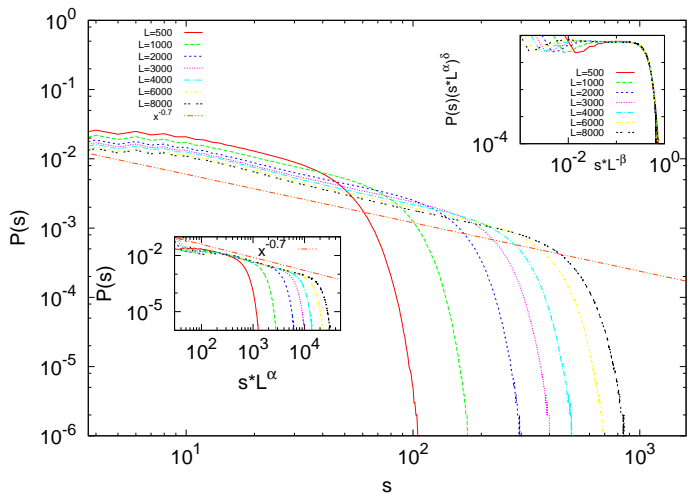}
   \caption{The avalanche size distributions for one dimensional train model with regular spacings of the blocks (IC), are plotted for
different system sizes. The inset shows the data collapse with the finite size scaling form assumed (Eq. \ref{scaling}). The exponent values 
are $\delta=0.70 \pm 0.01$, which is the equivalent of the Guttenberg-Richter law here, and the other scaling exponents are
$\alpha=0.40\pm 0.05$, $\beta=z-\alpha=0.80\pm 0.01$.}
\label{gr-regular}
\end{figure}

\noindent 
When the chain is pulled from one side, after some transients, the whole chain starts moving. All
statistics are to be taken once the whole chain starts moving.
 When an instability sets in (a slip event) the pulling 
is stopped. Successive slips may continue due to the first slip. We count the number of slip events on 
every scan of the lattice (with parallel updating). This number gives the
size $s$ of the avalanche and the time is increased by unity after each scan of the blocks of the upper chain. 
Fig. \ref{gr-regular} shows the avalanche size distributions for one dimensional IC model.
The scaling form assumed here (for this and the subsequent versions of this model) is the following
\begin{equation}
P(s) \sim \left(sL^{\alpha}\right)^{-\delta}f\left(\frac{sL^{\alpha}}{L^z}\right),
\label{scaling}
\end{equation}
where $P(s)$ is the probability of an avalanche of size $s$. The scaling relation tells us that at constant time 
$t$, the average avalanche size $\langle s\rangle$ decreases with $L$ as $L^{-\alpha}$. The values of the exponent obtained here are: 
$\delta=0.70\pm 0.01$, $\alpha=0.40 \pm 0.05$
and $z=1.20 \pm 0.06$. These values do not depend on the distribution of the pinning force.
Note that the exponent values are in contrast with those obtained before \cite{chianca}. However,
with the earlier definition of avalanches we get back $\delta \sim 1.5$, with no finite size scalings.
%%%%%%%%%%%%%%%%%%%%%%%%%%%%%%%%%%%%%%%%%%%%%%%%%%%%%%%%%%%%%%%%%%%%%%%%%%%%%%%%%%%%%%%%%%%%%%%%%%%%%%%%%%%%%%%%%%%%%%%%%%%%%%%%%%%%%%%%%
%%%%%%%%%%%%%%%%%%%%%%%%%%%%%%%%%%%%%%%%%%%%%%%%%%%%%%%%%%%%%%%%%%%%%%%%%%%%%%%%%%%%%%%%%%%%%%%%%%%%%%%%%%%%%%%%%%%%%%%%%%%%%%%%%%%%%%%%%%
%%%%%%%%%%%%%%%%%%%%%%%%%%%%%%%%%%%%%%%%%%%%%%%%%%%%%%%%%%%%%%%%%%%%%%%%%%%%%%%%%%%%%%%%%%%%%%%%%%%%%%%%%%%%%%%%%%%%%%%%%%%%%%%%%%%%%%%%%
 
\begin{figure}[tb]
\centering \includegraphics[height=6.7cm]{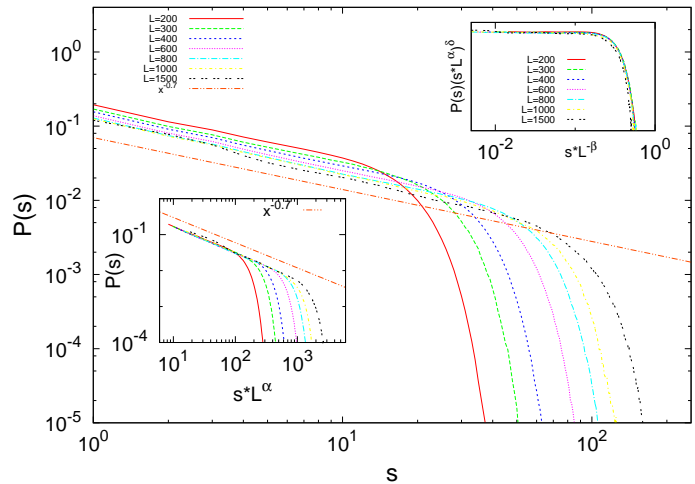}
   \caption{The avalanche size distributions for one dimensional train model with a fraction ($0.5$) of blocks removed from
 both upper and lower chains (DC), are plotted for
different system sizes. The inset shows the data collapse with the finite size scaling form assumed (Eq. \ref{scaling}). The exponent values 
are $\delta=0.70 \pm 0.01$, which is the equivalent of the Guttenberg-Richter law here, and the other scaling exponents are
$\alpha=0.40\pm 0.05$, $\beta=z-\alpha=0.80\pm 0.01$.}
\label{gr-random}
\end{figure}

In Fig. \ref{gr-random} we show the distribution of avalanche sizes when a  fraction ($0.5$) of blocks are
randomly removed from the upper and lower chains (disordered chain: DC). We do a similar finite size 
scaling study as we did for IC. The exponent values come out
to be very close to those obtained before. We have checked that the values do not depend on the fraction of the blocks removed, or
even for different distribution functions for the inter-block spacings (for random removal it will be exponential; we have checked for uniform, power-law
distributions etc.).  

\begin{figure}[tb]
\centering \includegraphics[height=6.7cm]{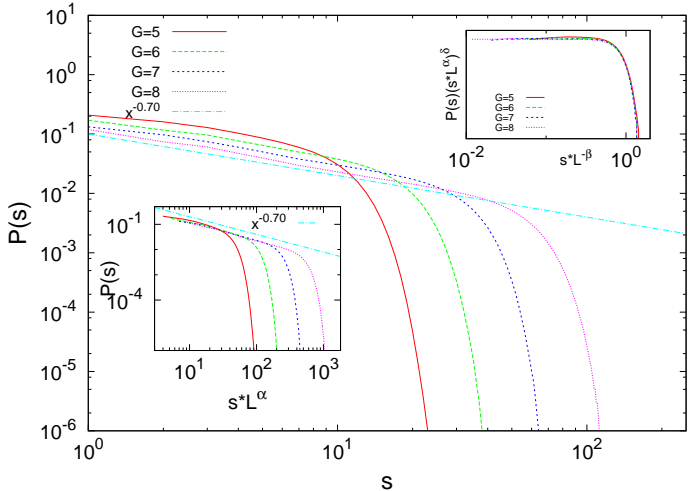}
   \caption{The avalanche size distributions for one dimensional train model with the blocks arranged in form of Cantor sets, are plotted for
different generation number. The inset shows the data collapse with the finite size scaling form assumed (Eq. \ref{scaling}). The exponent values 
are $\delta=0.70 \pm 0.01$, which is the equivalent of the Guttenberg-Richter law here, and the other scaling exponents are
$\alpha=0.40\pm 0.05$, $\beta=z-\alpha=0.80\pm 0.01$.}
\label{gr-fractal}
\end{figure}

Next, we consider the particular case when the disorders in the two chains are self-similar and 
arranged in the form of Cantor sets of some finite generation. Rough surfaces were modelled with 
Cantor sets before in the context of earthquakes \cite{tfo,book}. That the distribution of friction 
values becomes non-trivial even though a single pinning value is taken in every
contact is also known \cite{ebc} (see also \cite{penna} for stochastic friction coefficient). 
The earlier studies are infact the limiting cases of the present one 
when the springs are absolutely rigid. The avalanche statistics 
is plotted in Figs. \ref{gr-fractal}. Note that
while constructing the Cantor set, in each generation we have removed the middle third of the chain. So, while performing the finite size
scaling, the system size used was $2^G$, where $G$ is the generation number. We have considered a single 
value for the pinning force as in the DC case.
The exponent values obtained remain similar to the values obtained for IC and DC cases. 
 In this sense, the avalanche statistics are universal.
%%%%%%%%%%%%%%%%%%%%%%%%%%%%%%%%%%%%%%%%%%%%%%%%%%%%%%%%%%%%%%%%%%%%%%%%%%%%%%%%%%%%%%%%%%%%%%%%%%%%%%%%%%%%%%%%%%%%%%%%%%%%%%%%%%%%%%%%%

\begin{figure}[tb]
\centering \includegraphics[width=8.2cm]{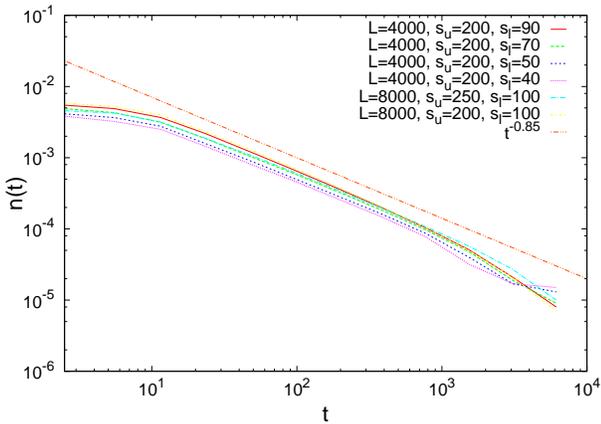}
   \caption{The probability that an avalanche of size $s_{l}$ or above takes places after a time $t$ of a main-shock (an event of size equal to or higher than $s_{u}$) is plotted for IC. This shows a power-law dependence and the exponent value is $0.85 \pm 0.05$. As can be seen, this value is insensitive to the cut-off values imposed. The power-law dependence is analogous to the Omori-law.}
\label{omori-regular}
\end{figure}

\begin{figure}[tb]
\centering \includegraphics[width=8.2cm]{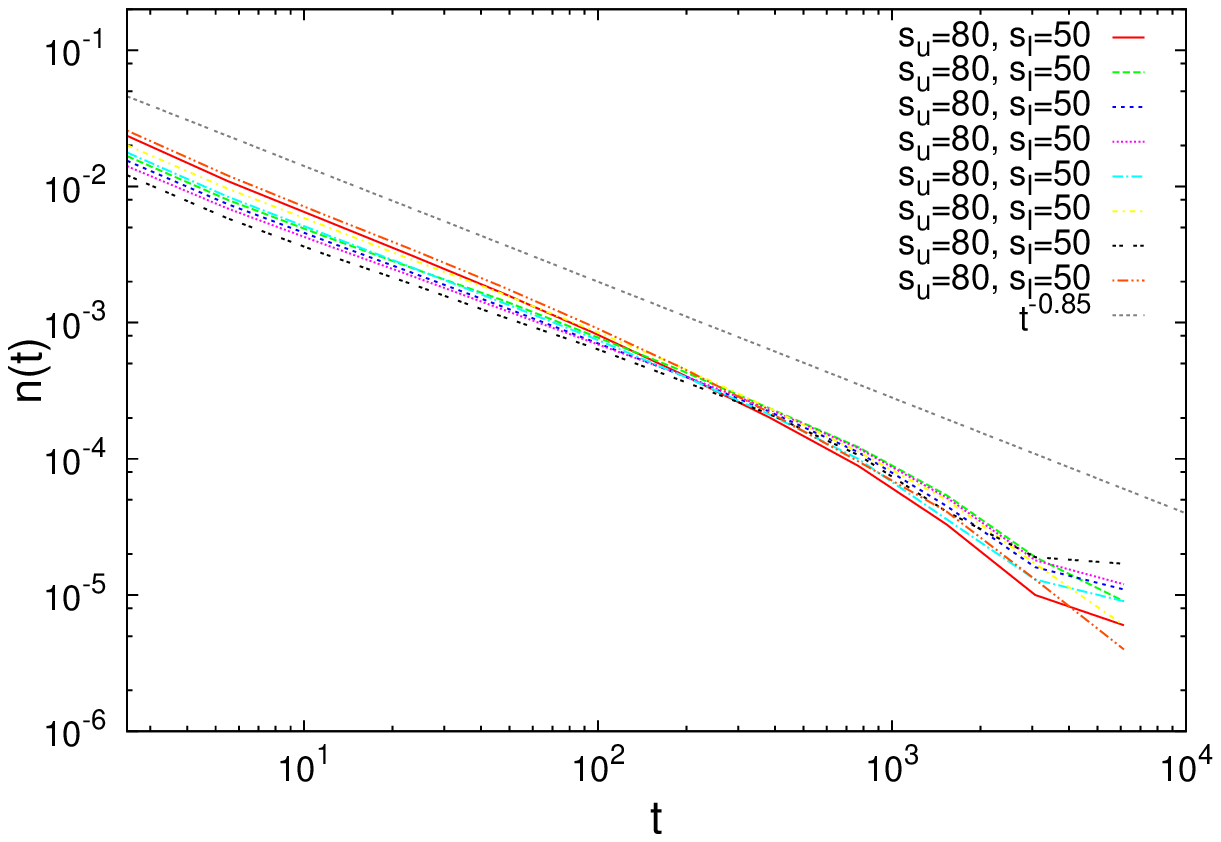}
   \caption{The probability that an avalanche of size $s_{l}$ or above takes places after a time $t$ of a main-shock (an event of size 
equal to or higher than $s_{u}$) is plotted for DC. This shows a power-law dependence and the exponent value is $0.85 \pm 0.05$. As can be seen,
this value is slightly sensitive with the cut-off values imposed externally. This power-law dependence is analogous to the Omori-law.}
\label{omori-random}
\end{figure}

\begin{figure}[tb]
\centering \includegraphics[width=8.2cm]{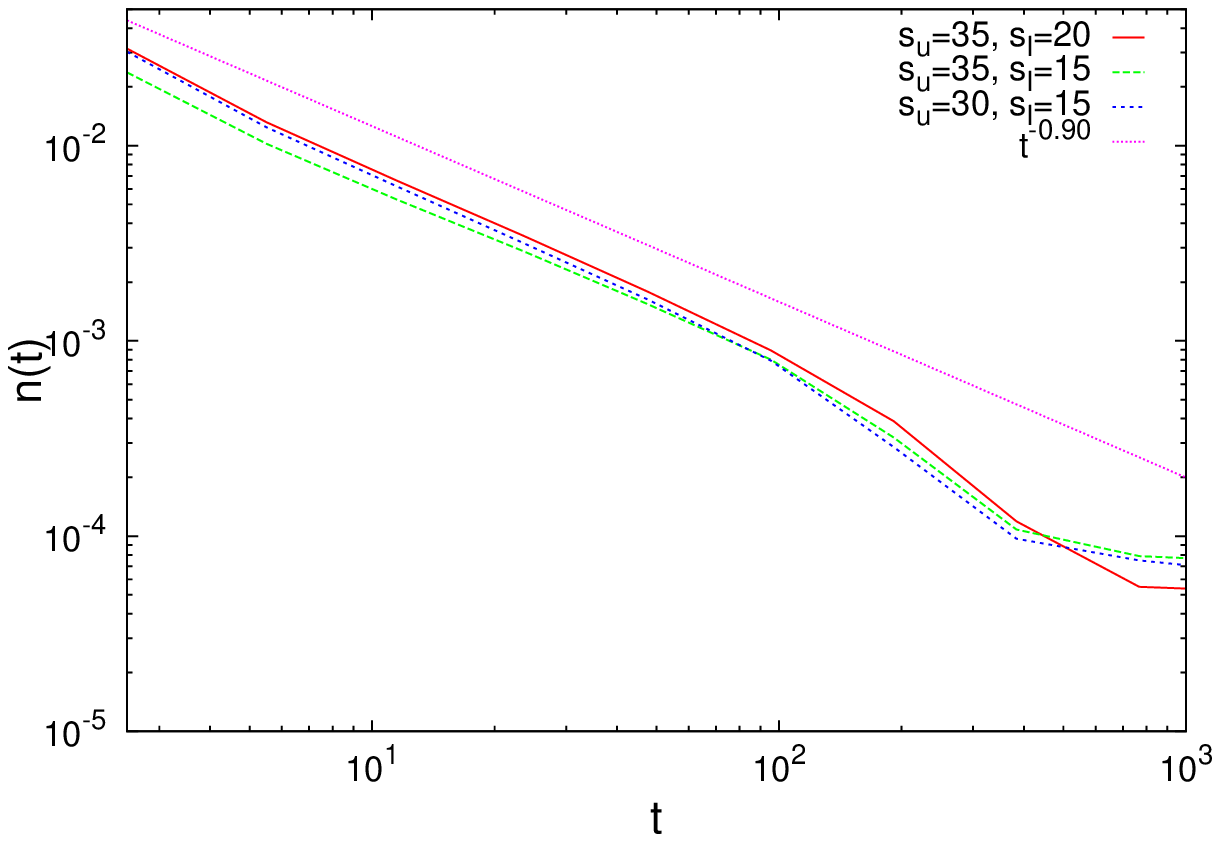}
   \caption{The probability that an avalanche of size $s_{l}$ or above takes places after a time $t$ of a main-shock (an event of size 
equal to or higher than $s_{u}$) is plotted for SAC. This shows a power-law dependence and the exponent value is $0.85 \pm 0.05$. As can be seen,
this value is slightly sensitive with the cut-off values imposed externally. This power-law dependence is analogous to the Omori-law.}
\label{omori-fractal}
\end{figure}

Next we intend to study the analogue of Omori law in this context. We first define an upper 
cut-off ($s_{u}$) of the avalanche size. The avalanches above this size is to be called a main-shock. 
We also define a lower cut-off ($s_{l}$), below which we do not measure an avalanche. We 
measure the probability that an avalanche of size $s_{l}$ or above has occurred after time $t$ after a 
main-shock. This is found to decay as a power-law
with time 
\begin{equation}
n(t)\sim t^{-p}
\label{omori}
\end{equation}
when $t$ is large  and the exponent value $p=0.85\pm 0.05$ is not sensitive to the cut-off values (See Fig. \ref{omori-regular} for IC, Fig.\ref{omori-random} for DC models and Fig.\ref{omori-fractal} for SAC model). 
%%%%%%%%%%%%%%%%%%%%%%%%%%%%%%%%%%%%%%%%%%%%%%%%%%%%%%%%%%%%%%%%%%%%%%%%%%%%%%%%%%%%%%%%%%%%%%%%%%%%%%%%%%%%%%%%%%%%%%%%%%%%%%%%%%%%%%%%%
%%%%%%%%%%%%%%%%%%%%%%%%%%%%%%%%%%%%%%%%%%%%%%%%%%%%%%%%%%%%%%%%%%%%%%%%%%%%%%%%%%%%%%%%%%%%%%%%%%%%%%%%%%%%
%%%%%%%%%%%%%%%%%%%%%%%%%%%%%%%%%%%%%%%%%%%%%%%%%%%%%%%%%%%%%%%%%%%%%%%%%%%%%%%%%%%%%%%%%%%%%%%%%%%%%%%%%%%%%%%%%%%%%%%%%%%%%%%%%%%%%%%%%
%%%%%%%%%%%%%%%%%%%%%%%%%%%%%%%%%%%%%%%%%%%%%%%%%%%%%%%%%%%%%%%%%%%%%%%%%%%%%%%%%%%%%%%%%%%%%%%%%%%%%%%%%%%%%%%%%%%%%%%%%%%%%%%%%%%%%%%%%%
%%%%%%%%%%%%%%%%%%%%%%%%%%%%%%%%%%%%%%%%%%%%%%%%%%%%%%%%%%%%%%%%%%%%%%%%%%%%%%%%%%%%%%%%%%%%%%%%%%%%%%%%%%%%%%%%%%%%%%%%%%%%%%%%%%%%%%%%%%%%%%%%%%%%%%%%%%%%%%%%%%%%%%%%%%%%%%%%%%%%%%%%%%%%%%%%%
%%%%%%%%%%%%%%%%%%%%%%%%%%%%%%%%%%%%%%%%%%%%%%%%%%%%%%%%%%%%%%%%%%%%%%%%%%%%%%%%%%%%%%%%%%%%%%%%%%%%%%%%%%%%%%%%%%%%%%%%%%%%%%%%%%%%%%%%%%
%%%%%%%%%%%%%%%%%%%%%%%%%%%%%%%%%%%%%%%%%%%%%%%%%%%%%%%%%%%%%%%%%%%%%%%%%%%%%%%%%%%%%%%%%%%%%%%%%%%%%%%%%%%%%%%%%%%%%%%%%%%%%%%%%%%%%%%%%
\begin{figure}[tb]
\centering \includegraphics[height=6.7cm]{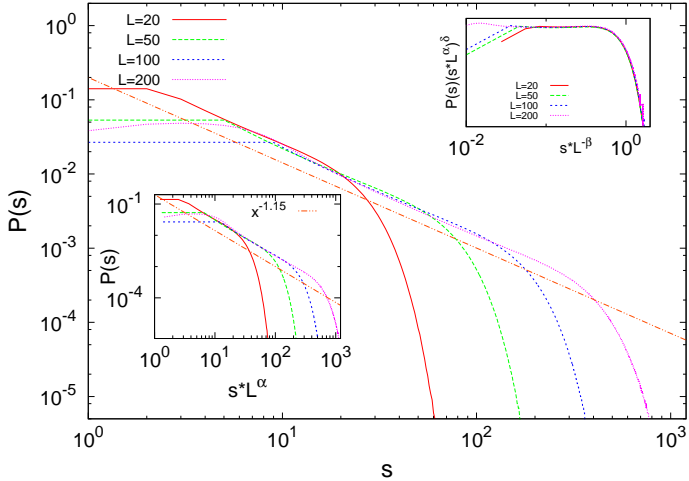}
   \caption{The avalanche size distributions for  train model with the blocks arranged in form of two dimensional lattice, are plotted for
different system sizes. The inset shows the data collapse with the finite size scaling form assumed (Eq. \ref{scaling}). The exponent values 
are $\delta=1.15 \pm 0.01$, which is the equivalent of the Guttenberg-Richter law here, and the other scaling exponents are
$\alpha=0.07\pm 0.01$, $\beta=z-\alpha=1.20\pm 0.01$.}
\label{gr-regular-2d}
\end{figure}
%%%%%%%%%%%%%%%%%%%%%%%%%%%%%%%%%%%%%%%%%%%%%%%%%%%%%%%%%%%%%%%%%%%%%%%%%%%%%%%%%%%%%%%%%%%%%%%%%%%%%%%%%%%%%%%%%%%%%%%%%%%%%%%%%%%%%%%%%%
%%%%%%%%%%%%%%%%%%%%%%%%%%%%%%%%%%%%%%%%%%%%%%%%%%%%%%%%%%%%%%%%%%%%%%%%%%%%%%%%%%%%%%%%%%%%%%%%%%%%%%%%%%%%%%%%%%%%%%%%%%%%%%%%%%%%%%%%%
%%%%%%%%%%%%%%%%%%%%%%%%%%%%%%%%%%%%%%%%%%%%%%%%%%%%%%%%%%%%%%%%%%%%%%%%%%%%%%%%%%%%%%%%%%%%%%%%%%%%%%%%%%%%%%%%%%%%%%%%%%%%%%%%%%%%%%%%%
\begin{figure}[tb]
\centering \includegraphics[width=8.2cm]{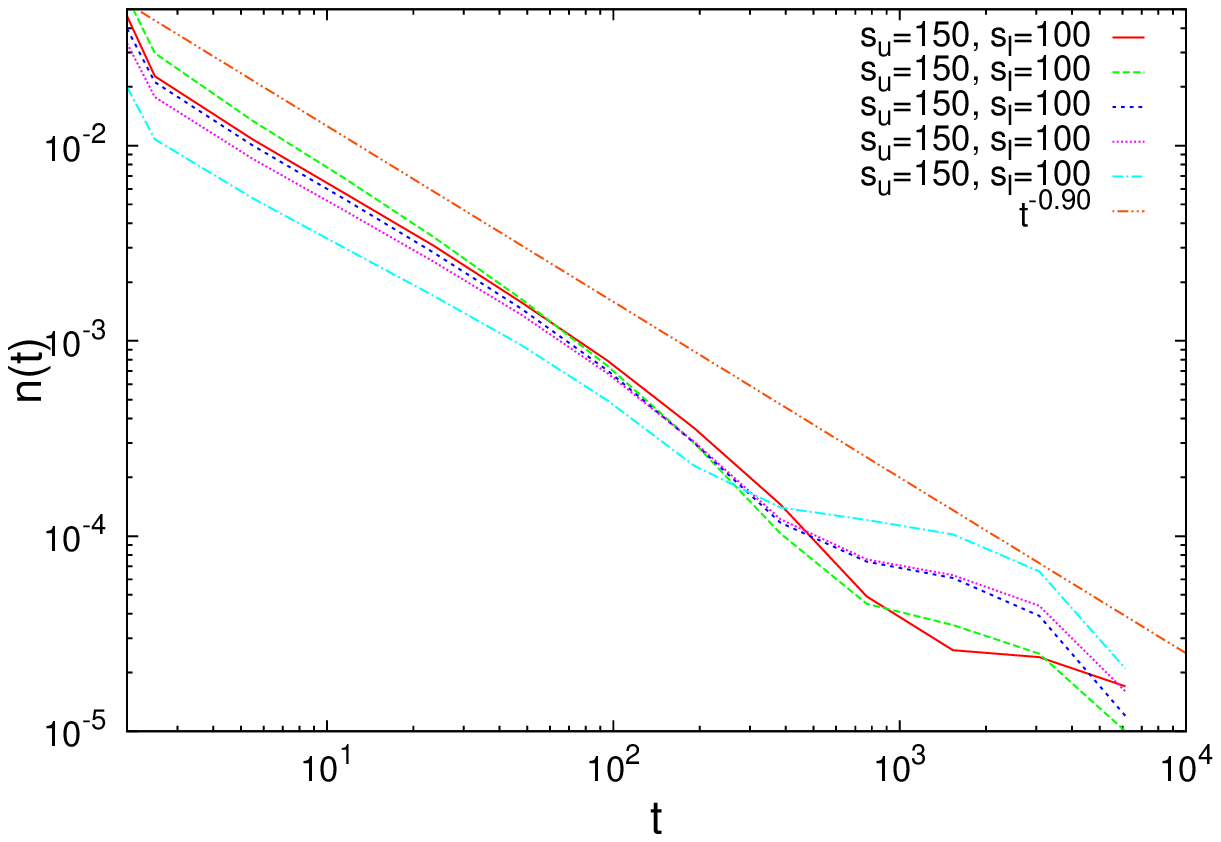}
   \caption{The probability that an avalanche of size $s_{l}$ or above takes places after a time $t$ of a main-shock (an event of size 
equal to or higher than $s_{u}$) is plotted for two dimensional train model. This shows a power-law dependence and the exponent value is $0.90 \pm 0.05$.}
\label{omori-regular-2d}
\end{figure}
%%%%%%%%%%%%%%%%%%%%%%%%%%%%%%%%%%%%%%%%%%%%%%%%%%%%%%%%%%%%%%%%%%%%%%%%%%%%%%%%%%%%%%%%%%%%%%%%%%%%%%%%%%%%%%%%%%%%%%%%%%%%%%%%%%%%%%%%%%
%%%%%%%%%%%%%%%%%%%%%%%%%%%%%%%%%%%%%%%%%%%%%%%%%%%%%%%%%%%%%%%%%%%%%%%%%%%%%%%%%%%%%%%%%%%%%%%%%%%%%%%%%%%%%%%%%%%%%%%%%%%%%%%%%%%%%%%%%
%%%%%%%%%%%%%%%%%%%%%%%%%%%%%%%%%%%%%%%%%%%%%%%%%%%%%%%%%%%%%%%%%%%%

Next, we consider the model in two dimension. The blocks are initially arranged in a regular square 
lattice ($L\times L$) and the system is pulled from one side (all $L$ blocks) quasistatically. 
The definition of avalanches remain the same. We consider the case of random 
pinning force distributed uniformly and the lattice is regular (IC). The exponent values for the avalanche statistics are very different that those for the one dimensional case.
As can be seen from Fig. \ref{gr-regular-2d}, the exponent values for 2d becomes $\alpha=0.07\pm 0.01$, $\delta=1.15\pm 0.02$ and $\beta=1.20\pm 0.01$.
However, the analogue of Omori-law (see Fig. \ref{omori-regular-2d}) remains almost similar, although the power-law fit is not very good in this case.

\subsection{Equivalence with interface depinning}
\noindent The avalanche statistics of the discrete train model studied here is similar to the statistics of the
interface depinning problem. Particularly, this linear elastic model with threshold activated dynamics is formally
similar to the Edwards-Wilkinson (EW) equation with quenched randomness \cite{ew}. The EW equation for interface
depinning reads
\begin{equation}
\frac{\partial h(x,t)}{\partial t}=\nabla ^2(h(x,t))+\eta(h(x,t))+f_{ext} \nonumber
\end{equation} 
where $h(x,t)$ is the height of the interface (from some arbitrary reference) on position $x$ at time $t$, $\eta$ is the quenched noise
and $f_{ext}$ is the applied force, which is uniform at all points. This equation shows a depinning transition in the sense
that if one measures the steady state velocity of the interface for different values of external force, one gets it in the form
\begin{eqnarray}
v_{sat} &\sim & (f_{ext}-f_c)^{\theta} \hskip0.5cm \mbox{ when} \hskip0.5cm f_{ext}>f_c \nonumber \\
v_{sat} & = & 0 \hskip0.5cm \mbox{otherwise.} 
\end{eqnarray}
where $\theta$ is the velocity depinning exponent. Depinning transition in bead spring model was studied before \cite{cule}.
However, that model was not discretised and the driving was applied on each bead, where Omori law like features cannot be observed.
%%%%%%%%%%%%%%%%%%%%%%%%%%%%%%%%%%%%%%%%%%%%%%%%%%%%%%%%%%%%%%%%%%%%
%%%%%%%%%%%%%%%%%%%%%%%%%%%%%%%%%%%%%%%%%%%%%%%%%%%%%%%%%%%%%%%%%%%%%%%%%%%%%%%%%%%%%%%%%%%%%%%%%%%%%%%%%%%%%%%%%%%%%%%%%%%%%%%%%%%%%%%%%
\begin{figure}[tb]
\centering \includegraphics[height=6.7cm]{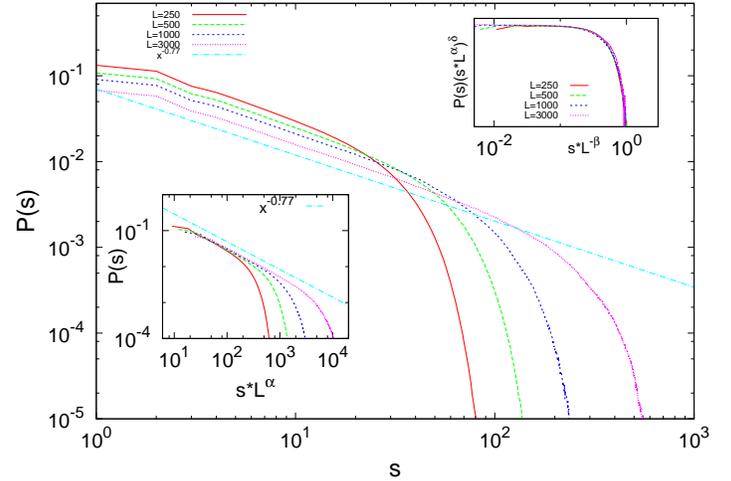}
   \caption{The avalanche size distributions for one dimensional EW model are plotted for
different system sizes. The inset shows the data collapse with the finite size scaling form assumed (Eq. \ref{scaling}). The exponent values 
are $\delta=0.77 \pm 0.02$, which is the equivalent of the Guttenberg-Richter law here, and the other scaling exponents are
$\alpha=0.40\pm 0.01$, $\beta=z-\alpha=0.82\pm 0.01$.}
\label{gr-ew-1d}
\end{figure}
%%%%%%%%%%%%%%%%%%%%%%%%%%%%%%%%%%%%%%%%%%%%%%%%%%%%%%%%%%%%%%%%%%%%%%%%%%%%%%%%%%%%%%%%%%%%%%%%%%%%%%%%%%%%%%%%%%%%%%%%%%%%%%%%%%%%%%%%%%
%%%%%%%%%%%%%%%%%%%%%%%%%%%%%%%%%%%%%%%%%%%%%%%%%%%%%%%%%%%%%%%%%%%%%%%%%%%%%%%%%%%%%%%%%%%%%%%%%%%%%%%%%%%%%%%%%%%%%%%%%%%%%%%%%%%%%%%%%
%%%%%%%%%%%%%%%%%%%%%%%%%%%%%%%%%%%%%%%%%%%%%%%%%%%%%%%%%%%%%%%%%%%%%%%%%%%%%%%%%%%%%%%%%%%%%%%%%%%%%%%%%%%%%%%%%%%%%%%%%%%%%%%%%%%%%%%%%
\begin{figure}[tb]
\centering \includegraphics[width=8.2cm]{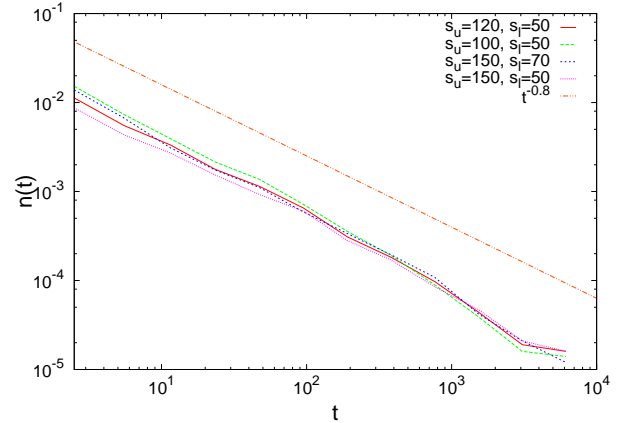}
   \caption{The probability that an avalanche of size $s_{l}$ or above takes places after a time $t$ of a main-shock (an event of size 
equal to or higher than $s_{u}$) is plotted for one dimensional EW model. This shows a power-law dependence and the exponent value is $0.80 \pm 0.05$.}
\label{omori-ew-1d}
\end{figure}
%%%%%%%%%%%%%%%%%%%%%%%%%%%%%%%%%%%%%%%%%%%%%%%%%%%%%%%%%%%%%%%%%%%%%%%%%%%%%%%%%%%%%%%%%%%%%%%%%%%%%%%%%%%%%%%%%%%%%%%%%%%%%%%%%%%%%%%%%%
%%%%%%%%%%%%%%%%%%%%%%%%%%%%%%%%%%%%%%%%%%%%%%%%%%%%%%%%%%%%%%%%%%%%%%%%%%%%%%%%%%%%%%%%%%%%%%%%%%%%%%%%%%%%%%%%%%%%%%%%%%%%%%%%%%%%%%%%%
%%%%%%%%%%%%%%%%%%%%%%%%%%%%%%%%%%%%%%%%%%%%%%%%%%%%%%%%%%%%%%%%%%%%
\noindent Apart from this steady state exponent $\theta$, one can also study the avalanche dynamics in this model when it is driven 
quasistatically at the boundaries. That the boundary driven interface dynamics and train model can be in the same
universality class was conjectured before \cite{maya} (see also \cite{sor}).  Here, we show that the discrete version of the train model with random pinning forces is exclusively the boundary driven interface model.

\begin{figure}[ht]
\begin{center}
\includegraphics[scale=0.65]{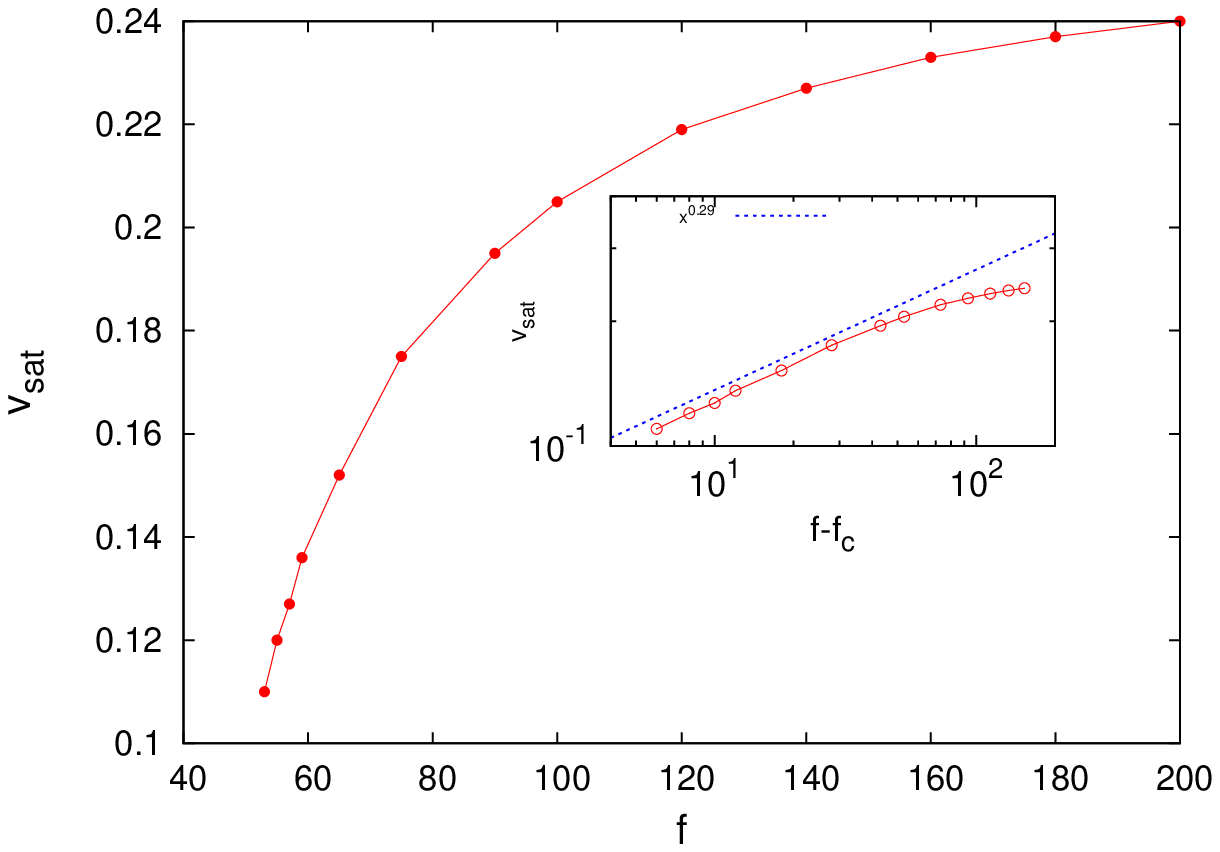}
\includegraphics[scale=0.65]{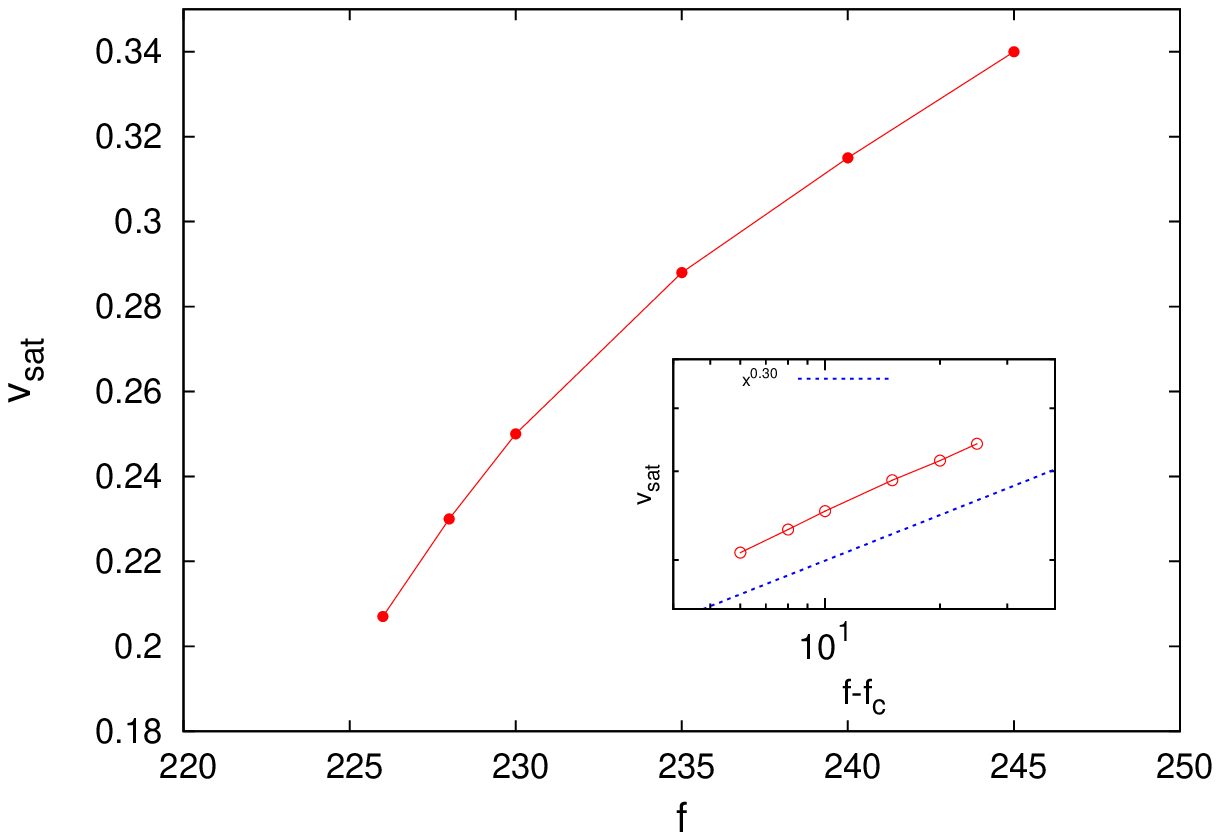}
\end{center}
\caption{The steady state velocities as a function of external force for train model (top) and EW model (bottom) 
in one dimension. In both cases the velocity increases in a power-law beyond a critical point. The exponent are $0.29\pm 0.01$
and $0.30 \pm 0.01$ for train model and boundary driven EW model respectively.}
\label{vel-sat}
\end{figure}

The avalanche statistics, with avalanches defined as before, in the interface problem is same as in the case of
train model (in 1d). In Fig. \ref{gr-ew-1d} , we plot the avalanche size distribution and its finite size scaling and get the same exponents.
We also measure the rate of events after a large shock and get similar statistics (see Fig. \ref{omori-ew-1d}). As mentioned before, 
we also measure the steady state order parameter exponent $\theta$ both in train model and EW model. The results are
shown in Fig. \ref{vel-sat}. The proximity in the values of the exponents suggest that these are manifestations of the same model.
%%%%%%%%%%%%%%%%%%%%%%%%%%%%%%%%%%%%%%%%%%%%%%%%%%%%%%%%%%%%%%%%%%%%%%%%%%%%%%%%%%%%%%%%%%%%%%%%%%
%%%%%%%%%%%%%%%%%%%%%%%%%%%%%%%%%%%%%%%%%%%%%%%%%%%%%%%%%%%%%%%%%%%%%%%%%%%%%%%%%%%%%%%%%%%%%%%%
%%%%%%%%%%%%%%%%%%%%%%%%%%%%%%%%%%%%%%%%%%%%%%%%%%%%%%%%%%%%%%%%%%%%%%%%%%%%%%%%%%%%%%%%%%%%
%%%%%%%%%%%%%%%%%%%%%%%%%%%%%%%%%%%%%%%%%%%%%%%%%%%%%%%%%%%%%%%%%%%%%%%%%%%%%%%%%%%%%%%%%%%
%%%%%%%%%%%%%%%%%%%%%%%%%%%%%%%%%%%%%%%%%%%%%%%%%%%%%%%%%%%%%%%%%%%%
%%%%%%%%%%%%%%%%%%%%%%%%%%%%%%%%%%%%%%%%%%%%%%%%%%%%%%%%%%%%%%%%%%%%%%%%%%%%%%%%%%%%%%%%%%%%%%%%%%%%%%%%%%%%%%%%%%%%%%%%%%%%%%%%%%%%%%%%%
\begin{figure}[tb]
\centering \includegraphics[height=6.7cm]{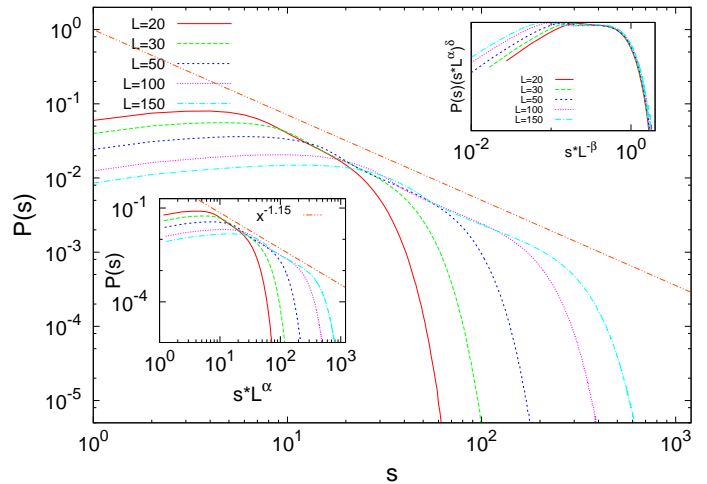}
   \caption{The avalanche size distributions for two dimensional EW model (driven at the boundaries) are plotted for
different system sizes. The inset shows the data collapse with the finite size scaling form assumed (Eq. \ref{scaling}). The exponent values 
are $\delta=1.15 \pm 0.01$, which is the equivalent of the Guttenberg-Richter law here, and the other scaling exponents are
$\alpha=0.05\pm 0.01$, $\beta=z-\alpha=1.20\pm 0.01$.}
\label{gr-ew-2d}
\end{figure}
%%%%%%%%%%%%%%%%%%%%%%%%%%%%%%%%%%%%%%%%%%%%%%%%%%%%%%%%%%%%%%%%%%%%%%%%%%%%%%%%%%%%%%%%%%%%%%%%%%%%%%%%%%%%%%%%%%%%%%%%%%%%%%%%%%%%%%%%%%
%%%%%%%%%%%%%%%%%%%%%%%%%%%%%%%%%%%%%%%%%%%%%%%%%%%%%%%%%%%%%%%%%%%%%%%%%%%%%%%%%%%%%%%%%%%%%%%%%%%%%%%%%%%%%%%%%%%%%%%%%%%%%%%%%%%%%%%%%
%%%%%%%%%%%%%%%%%%%%%%%%%%%%%%%%%%%%%%%%%%%%%%%%%%%%%%%%%%%%%%%%%%%%%%%%%%%%%%%%%%%%%%%%%%%%%%%%%%%%%%%%%%%%%%%%%%%%%%%%%%%%%%%%%%%%%%%%%
\begin{figure}[tb]
\centering \includegraphics[width=8.2cm]{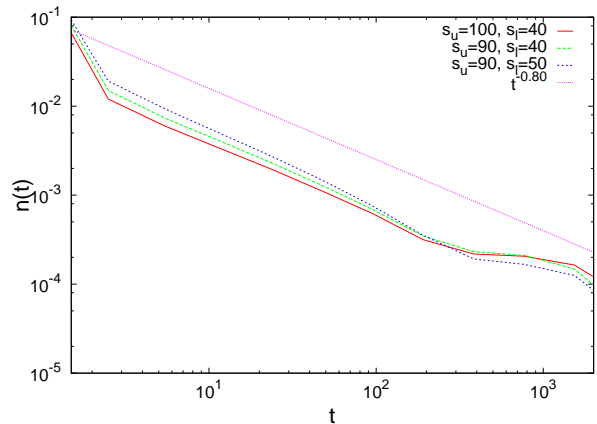}
   \caption{The probability that an avalanche of size $s_{l}$ or above takes places after a time $t$ of a main-shock (an event of size 
equal to or higher than $s_{u}$) is plotted for two dimensional EW model. This shows a power-law dependence and the exponent value is $0.80 \pm 0.02$.}
\label{omori-ew-2d}
\end{figure}
%%%%%%%%%%%%%%%%%%%%%%%%%%%%%%%%%%%%%%%%%%%%%%%%%%%%%%%%%%%%%%%%%%%%%%%%%%%%%%%%%%%%%%%%%%%%%%%%%%%%%%%%%%%%%%%%%%%%%%%%%%%%%%%%%%%%%%%%%%
%%%%%%%%%%%%%%%%%%%%%%%%%%%%%%%%%%%%%%%%%%%%%%%%%%%%%%%%%%%%%%%%%%%%%%%%%%%%%%%%%%%%%%%%%%%%%%%%%%%%%%%%%%%%%%%%%%%%%%%%%%%%%%%%%%%%%%%%
Furthermore, we study EW model in 2d to compare with train model in 2d. Here we drive the 2d surface again from the boundaries.
The avalanche statistics and rate of activities are plotted in Figs. \ref{gr-ew-2d}, \ref{omori-ew-2d}. Again these are similar 
to the values obtained in 
train model. Therefore we see that the steady state and dynamical behaviours of the discrete train model with 
random pinning are same as that of EW model. In both cases the systems are boundary driven and are of linear elastic nature with
random pinning. 
\subsection{Interface propagation and fluctuation in bulk}
\begin{figure}[tb]
\centering \includegraphics[width=8.2cm]{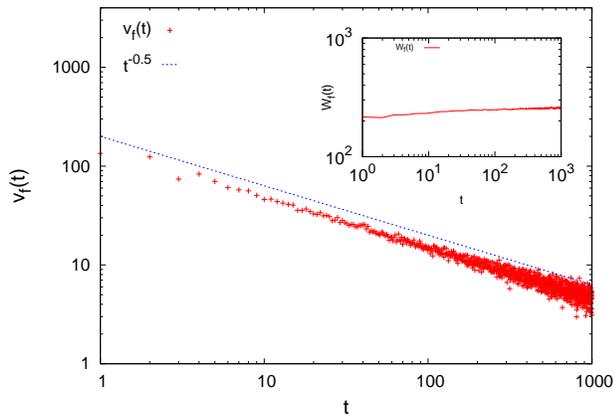}
   \caption{The velocity of the front $v(t)$ (upto which disturbance has propagated) is plotted against time for 2d EW model, showing
a $t^{-0.5}$ dependence. The inset shows the width of this front, which is practically independent of time (i.e., logarithmic
dependence).}
\label{ew-2d-ff}
\end{figure}

\begin{figure}[tb]
\centering \includegraphics[width=8.2cm]{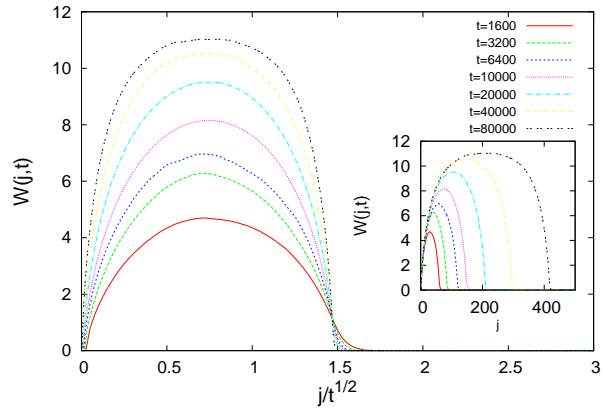}
   \caption{The fluctuation within the bulk of the system is measured along the direction perpendicular to the drive in 2d EW model.
After scaling the displacements, one finds a symmetric profile of fluctuation withing the bulk.}
\label{ew-2d-wei}
\end{figure}

\noindent As mentioned above, the drive in 2d EW model (and also in 2d train model) is applied along one side.
Therefore one may ask how the disturbances propagate through the bulk. Considering the case of 2d EW model (of course
everything can be translated to the train model in 2d as well), we study mainly two aspects of this disturbance
propagation: (a) Since the drive is applied only along a side, in the early time dynamics, there will be a `front' similar
to the propagation of interface through quenched disordered medium. This is a rough line with an average velocity of propagation
that decays in a power-law. In Fig. \ref{ew-2d-ff} the front velocity and the r.m.s. fluctuation of the front are plotted with
time. While the velocity decays as $v(t)\sim t^{-0.5}$, the front width does not show appreciable dependence with time, 
making it similar to a logarithmic growth. (b) We also measure the fluctuation of the system along the direction perpendicular 
to the drive. As one goes away from the line which is driven, the widths of  each line first increases, reaches a maximum and
then decreases to zero in the region which are yet to experience the drive. Since the front velocity scales as $t^{-0.5}$, the 
displacement (i.e., the maximum length upto which disturbance has propagated and one gets finite width) scales as $t^{0.5}$. 
One can therefore scale the displacement axis by $t^{0.5}$ and make the end-points coincide (see Fig. \ref{ew-2d-wei}). One then finds
a symmetric curve, i.e. the fluctuation in the bulk is maximum at the halfway of the range upto which it has propagated.  

%%%%%%%%%%%%%%%%%%%%%%%%%%%%%%%%%%%%%%%%%%%%%%%%%%%%%%%%%%%%%%%%%%%%%%%%%%%%%%%%%%%%%%%%%%%%%%%%%%%%%%%%%%%%%%%%%%%%%%%%%%%%%%%%%%%%%%%%%
%%%%%%%%%%%%%%%%%%%%%%%%%%%%%%%%%%%%%%%%%%%%%%%%%%%%%%%%%%%%%%%%%%%%%%%%%%%%%%%%%%%%%%%%%%%%%%%%%%%%%%%%%%%%%%%%%%%%%%%%%%%%%%%%%%%%%%%%%%
%%%%%%%%%%%%%%%%%%%%%%%%%%%%%%%%%%%%%%%%%%%%%%%%%%%%%%%%%%%%%%%%%%%%%%%%%%%%%%%%%%%%%%%%%%%%%%%%%%%%%%%%%%%%%%%%%%%%%%%%%%%%%%%%%%%%%%%%%
%%%%%%%%%%%%%%%%%%%%%%%%%%%%%%%%%%%%%%%%%%%%%%%%%%%%%%%%%%%%%%%%%%%%%%%%%%%%%%%%%%%%%%%%%%%%%%%%%%%%%%%%%%%%%%%%%%%%%%%%%%%%%%%%%%%%%%%%%
%%%%%%%%%%%%%%%%%%%%%%%%%%%%%%%%%%%%%%%%%%%%%%%%%%%%%%%%%%%%%%%%%%%%%%%%%%%%%%%%%%%%%%%%%%%%%%%%%%%%%%%%%%%%%%%%%%%%%%%%%%%%%%%%%%%%%%%%%%
%%%%%%%%%%%%%%%%%%%%%%%%%%%%%%%%%%%%%%%%%%%%%%%%%%%%%%%%%%%%%%%%%%%%%%%%%%%%%%%%%%%%%%%%%%%%%%%%%%%%%%%%%%%%%%%%%%%%%%%%%%%%%%%%%%%%%%%%

There are two things to be mentioned here, first the `front' may have overhangs.
To deal with it, we are considering the surface which can be formed by over-estimating the overhangs (see \cite{overhang}). One can estimate
the effect of over-hangs in other ways, and it may differ substantially only near a dynamical transition (depinning), which we are
not studying here. Second, while one can study the time dependence by quasistatically increasing the displacement at one
end and wait for the system to come to a halt (that makes one time step), one can also make the entire displacement at one go
and one time step consists of one scan of the lattice. These two approaches are not same in terms of avalanche statistics,
but in fluctuation in the bulk and velocity of the front, these two turns out to be same. Hence we followed the later, which 
is computationally cheaper.

\section{Summary and discussion}
\noindent We  study a discretised version of the train model \cite{vieira} of earthquake dynamics. The non-linear
velocity weakening friction term is replaced by  (random) pinning force threshold and the movements of the blocks are discretised,
in the sense that one block can move one lattice constant at a given time. This allows us to model the system as two chains;
 one dragged over the other by pulling the upper chain from one end. The blocks in the upper chain, which is being 
dragged, are connected by 
linear springs with identical spring constants, while the blocks in the lower chain are fixed in position (see Fig. \ref{chain}).
The (friction)  pinning force  is non-zero only in places where two blocks come on top of one another. When both the
chains are intact (IC model; blocks are placed at equal distances initially), whenever one block comes on top of  another, a random pinning
force is drawn from a uniform distribution in [$0:1$]. In the case (DC model), both the chains are disordered, i.e., some  blocks are randomly
removed from both the chains, and the pinning (friction) force is taken to be simply a constant (unity) whenever one block comes over another. 
We make the connection that this version of train model is precisely same as the Edwards-Wilkinson's model 
for interface propagation through quenched disordered medium. This is apparent from Eq. (\ref{dyn2}), when the EW front is driven 
by its two ends in the transverse direction (perpendicular to the initial position of the chain). 
Both of these models have also been studied in two dimensions, where in train model the system is driven along the plane of the 
surface, for EW model it is driven perpendicular to the surface. We obtain equivalent results.
That these two models may be in the same universality
class, was conjectured before (see e.g., \cite{maya}), but here we show that the dynamics is these 
two models can be translated to be precisely the same
with all its qualitative features intact and identical complimentary features: Power law  growth of depinning
velocity in the train model and Omori law in EW model.

The dynamics of the train model follows Eqs. (\ref{dyn1}) and (\ref{dyn2}). 
The pulling of the upper chain is stopped whenever a slip event
occurs. An avalanche is defined as the number of displacements in one scan of the (entire) upper chain (parallel updates are made).
 Time is increased by unity after 
each scan.  Further pulling is resumed 
only when all the blocks have relaxed. This makes an avalanche distinguishable from all the previously occurring avalanches
and allows for the study of aftershock statistics. 
The avalanche statistics changes significantly (see sec. III). 
In both the cases (IC and DC), the avalanche size distribution exponent becomes $\delta=0.70\pm0.01$ and $\alpha=0.40\pm0.01$, 
$z=1.20\pm0.01$ (Eq. \ref{scaling}).  The universality of the avalanche statistics  is clear: these exponent values remain unchanged 
irrespective of  the details of the models (intact chain, chain with random or Cantor set like disorder or values of spring constants etc.). 
Next we study the after-shock behaviour (Omori law) of the avalanches. We first search for upper size ($s_u$) of the
avalanche and choose a lower cut-off $s_l$. An 
avalanche of size greater than $s_u$ is then
taken as the main-shock. We calculate the probability of having an aftershock of size $s_l$ or above after time $t$ of the main-shock. 
This follows a scale free behavior (see Eq. (\ref{omori})) with exponent $0.85\pm 0.05$,  analogous to Omori law. 
This Omori law was not detected earlier 
in any spring-block type model. Like the EW model for depinning, we measure the depinned velocity for the train model (see Fig. \ref{vel-sat}).
The depinning exponent value ($\theta=0.29 \pm 0.01$) is the same as the EW depinning exponent value.
The exponent values mentioned above changes significantly in two dimensions (see Fig. \ref{gr-regular-2d}),
 though remain universal (independent of spring constant,
pinning force distribution etc.). The value of the avalanche exponent becomes $\delta=1.15 \pm 0.01$; the other finite size scaling
exponents are $\alpha=0.07\pm 0.01$, $z=1.27\pm 0.02$. The Omori law exponent value ($0.90\pm0.05$),
 however, remains close to one dimensional value.
The fluctuation propagation through the bulk in the 2d EW model was also studied. It is found that when
the disturbance (of pulling the system by one end line) has not reached the other end, the fluctuation is maximum
just half-way of the total distance (in units of lattice constants) the disturbance-front has propagated 
(see Fig. \ref{ew-2d-ff}). The disturbance-front has a power-law 
decay in velocity, with
exponent value $0.50\pm 0.01$ (see Fig. \ref{ew-2d-wei}), while the r.m.s. fluctuation grows only logarithmically.

In conclusion, a discretized version of the Burridge-Knopoff train model with (non-linear friction force replaced by)
 random pinning is studied in one and two dimensions. 
With suitable definitions of avalanches, a scale free distribution of avalanche and the Omori law type behaviour for
after-shocks are obtained.
With this simplification, the avalanche dynamics of the model becomes precisely similar (identical exponent values) 
to the Edwards-Wilkinson model of interface propagation.
It also allows the complimentary observation of depinning velocity growth (with exponent value  identical
with that for Edwards-Wilkinson model) in this train model and Omori law behaviour of after-shock (depinning) avalanches in the 
Edwards-Wilkinson model. The observations of universalities of avalanche dynamics and interface depinning dynamics should 
shed new light in the existing conjectures regarding the statistical similarity in the dynamics of interface and fracture \cite{fisher},
interface and earthquake \cite{maya} and the recent experimental observations regarding the dynamics of fracture and
earthquakes \cite{baro}.      
\vskip0.3cm
\thanks
\noindent The work was finalized during the visit of the authors to SINTEF, Norway. The authors acknowledge 
participants of the INDNOR Project meeting, particularly Srutarshi Pradhan, for useful discussions. The 
financial supports from  project INDNOR 217413/E20 (NFR, Govt. of Norway) and BKC’s JC Bose Fellowship (DST, Govt. of India) are 
acknowledged.
\vskip-.75cm
%
%
% BibTeX users please use
% \bibliographystyle{}
% \bibliography{}

\begin{thebibliography}{}
%
% and use \bibitem to create references.
%

\bibitem{bk}
R. Burridge, L. Knopoff, Bull. Seismol. Soc. A. {\bf 57} (1967) 341.


\bibitem{rmp2}
H. Kawamura, T. Hatano, N. Kato, S. Biswas, B. K. Chakrabarti, Rev. Mod. Phys. {\bf 84} (2012) 839.

\bibitem{vieira}
M. Vieira, Phys. Rev. A {\bf 46}, (1992) 6288.

\bibitem{chianca}
C. V. Chianca, J. S. S. Martins, P. M. C. de Oliveira, Eur. Phys. J. B {\bf 68} (2009) 549.



\bibitem{ew}
S. Edwards, D. Wilkinson, Proc. R. Soc. A {\bf 381} (1982) 17.

\bibitem{bouchaud}
D. Bonamy, E. Bouchaud, Phys. Rep. {\bf 498} (2011) 1;
%\bibitem{stanley}
{\it Fractal concepts in surface growth}, A.-L. Barab\'{a}si, H. E. Stanley, Cambridge University Press, Cambridge (1995).


\bibitem{tfo}
B. K. Chakrabarti, R. B. Stinchcombe, Physica A {\bf 270} (1999) 27; P. Bhattacharyya, Physica A {\bf 348} (2005) 199.

\bibitem{book}
{\it Modelling critical and catastrophic phenomena in geoscience}, P. Bhattacharyya, B. K. Chakrabarti (Eds.), Springer-Verlag,
Heidelberg (2006).

\bibitem{ebc}
J. A. Eriksen, S. Biswas, B. K. Chakrabarti, Phys. Rev. E {\bf 82}  (2010) 041124.

\bibitem{penna}
A. R. de Lima, C. F. Moukarzel, I. Grosse, T.J.P. Penna, Phys. Rev. E {\bf 61} (2000) 2267.

\bibitem{cule}
D. Cule, T. Hwa, Phys. Rev. Lett. {\bf 77} (1996) 278.

\bibitem{maya}
M. Paczuski, S. Boettcher, Phys. Rev. Lett {\bf 77} (1996) 111.

\bibitem{sor}
A. Malthe-S\o{}renssen, Phys. Rev. E {\bf 59} (1999) 4169.

\bibitem{overhang}
N. J. Zhou, B. Zheng, Phys. Rev. E {\bf 82}  (2010) 031139.

\bibitem{fisher}
S. Ramanathan, D. Fisher, Phys. Rev. B {\bf 58} (1998) 6026.

\bibitem{baro}
J. Bar\'{o}, A. Corral, X. Illa, A. Planes, E. K. H. Salije,
W. Schranz, D. E. Soto-Parra, E. Vives, Phys. Rev. Lett. {\bf 110} (2013) 088702. 




%\bibitem{RefJ}
% Format for Journal Reference
%Author, Journal \textbf{Volume}, (year) page numbers.
% Format for books
%\bibitem{RefB}
%Author, \textit{Book title} (Publisher, place year) page numbers
% etc
\end{thebibliography}
%
% Non-BibTeX users please use

\end{document}